\documentclass[pra,aps,twocolumn]{revtex4}
\usepackage{graphicx}

\begin{document}
\title{Polariton excitation in epsilon-near-zero slabs: transient trapping of slow light}

\author{Alessandro Ciattoni$^1$, Andrea Marini$^2$, Carlo Rizza$^{1,3}$, Michael Scalora$^4$ and Fabio Biancalana$^{2,5}$}
\affiliation{$^1$Consiglio Nazionale delle Ricerche, CNR-SPIN 67100 L'Aquila, Italy and Dipartimento di Fisica,
             Universit\`a dell'Aquila, 67100 L'Aquila, Italy}
\affiliation{$^2$Max Planck Institute for the Science of Light, Guenther-Scharowsky-Stra\ss e 1, 91058 Erlangen, Germany} \affiliation{$^3$Dipartimento di
Fisica e Matematica, Universit\`a dell'Insubria, Via Valleggio 11, 22100 Como, Italy} \affiliation{$^4$Charles M. Bowden Research Center RDMR-WDS-WO,
RDECOM, Redstone Arsenal, Alabama 35898-5000, USA} \affiliation{$^5$School of Engineering \& Physical Sciences, Heriot-Watt University, Edinburgh, EH14
4AS, United Kingdom}
\date{\today}

\begin{abstract}
We numerically investigate the propagation of a spatially localized and quasi-monochromatic electromagnetic pulse through a slab with Lorentz dielectric
response in the epsilon-near-zero regime, where the real part of the permittivity vanishes at the pulse carrier frequency. We show that the pulse is able
to excite a set of virtual polariton modes supported by the slab, the excitation undergoing a generally slow damping due to absorption and radiation
leakage. Our numerical and analytical approaches indicate that in its transient dynamics the electromagnetic field displays the very same enhancement of
the field component perpendicular to the slab, as in the monochromatic regime. The transient trapping is inherently accompanied by a significantly reduced
group velocity ensuing from the small dielectric permittivity, thus providing a novel platform for achieving control and manipulation of slow light.
\end{abstract}

\maketitle

\section{Introduction}

Physical mechanisms driving to slow and fast light have attracted considerable attention from the scientific community in the last decades
\cite{Tsakmakidis_NatLett_2007,Boyd_JModPhys_2009,Kim_OE_2012}. The inherent interest in slow light comes from the long matter-radiation interaction time,
which can lead to considerable enhancement of all nonlinear processes that in turn may be exploited for active functionalities \cite{Vlasov_Nature_2005},
e.g. all-optical switching and modulation \cite{Mingaleev_PRE_2006,Bajcsy_PRL_2009}. The nonlinearity may also be enhanced by reducing the effective area
in subwavelength silicon on insulator and plasmonic waveguides \cite{Afshar_OE_2009,Marini_PRA_2011}, where tight confinement opens up possibilities for
miniaturized nonlinear applications \cite{Koos_OE_2007,Palomba_PRL_2008,Kauranen_NatPhot_2012}. Alternatively, extreme nonlinear dynamics
\cite{Ciattoni_PRA_2010,Ciattoni_PRA_2011}, enhanced second and third harmonic generation \cite{Vincenti_PRA_2011,Ciattoni_PRA_2012} is predicted in
epsilon-near-zero (ENZ) metamaterials, where the linear susceptibility is tailored in such a way that its modulus becomes comparable with the nonlinear
counterpart. Boosting the nonlinearity of ENZ plasmonic channels can also lead to active control of tunneling \cite{Powell_PRB_2009}, switching and
bistable response \cite{Argyropoulos_PRBB_2012}. ENZ metamaterials have also been used for directive emission \cite{Enoch_PRL_2002,Alu_IEEE_2006},
cloaking \cite{Alu_PRE_2005}, energy squeezing in narrow channels \cite{Silveirinha_PRL_2006} and subwavelength imaging
\cite{Alu_PRB_2007,Castaldi_PRB_2012}.

In all of the above mentioned mechanisms, the trade-off needed to achieve enhanced active functionalities is paid in terms of increased losses. As a
result, the ENZ regime one usually invokes refers to the case where the real part of the susceptibility becomes very small, while its imaginary part
remains finite. Indeed, due to the stringent physical requirement of causality, Kramers-Kronig relations impose that dispersion be inherently accompanied
by loss and the dielectric susceptibility can not become rigorously null. The residual loss either limits or even prevents giant enhancement of coherent
mechanisms, e.g. in second and third harmonic generation setups \cite{Vincenti_PRA_2011,Ciattoni_PRA_2012}. Recently, in the context of surface plasmon
polaritons, a method has been proposed to overcome the loss barrier for superlensing applications by loading the effect of loss into the time domain
\cite{Archambault_PRL_2012}. In our analytical calculations, we will use a similar approach to study the behavior of an electromagnetic pulse that
scatters from a slab having a Lorentz dielectric response. Indeed, by considering non-monochromatic {\it virtual modes} with complex frequency
\cite{Archambault_PRB_2009}, it is possible to drop off the effect of loss on the temporal dependence of the ``mode'' itself. In this complex frequency
approach, it is possible to achieve the condition where the dielectric susceptibility exactly vanishes. Our formalism treats the dielectric polarization
of the medium as a generic Lorentz oscillator that, in the epsilon-equal-to-zero condition, encompasses longitudinal collective oscillations of both
electrons (volume plasmons) and ions (longitudinal phonons) that can not be excited by light \cite{Ashcroft_Book}. Recently, the question whether or not
volume plasmons can be excited or not by classical light has been revived
\cite{Henrard_SynthMet_1999,Hoeflich_PRL_2009,Henrard_PRL_2010,Hoeflich_PRL_2010,Muys_OL_2012}. Some studies on Mie extinction efficiencies reveal a
maximum around the characteristic frequency where the dielectric susceptibility vanishes, attributing the enhanced extinction to the excitation of volume
plasmons \cite{Hoeflich_PRL_2009,Hoeflich_PRL_2010}. Conversely, other similar studies identify the physical origin of the enhanced extinction in the
excitation of leaky modes \cite{Henrard_SynthMet_1999,Henrard_PRL_2010}. The latter interpretation is also supported by studies of the excitation of
surface phonon polaritons in ENZ slabs \cite{Berreman_PhysRev_1963,Ruppin_RepProgPhys_1970,Vassant_OE_2012,Vassant_PRL_2012}.

In this manuscript we numerically investigate and analytically interpret the scattering of a spatially and temporally localized optical pulse from a
dielectric slab in the ENZ regime. We used a finite difference time domain (FDTD) algorithm to solve the full vectorial Maxwell equations coupled to the
Lorentz oscillator equation for the dielectric polarization of the slab. We find that, if the carrier frequency of the optical pulse matches the ENZ
condition, electromagnetic quasi-trapping occurs within the Lorentz slab since, after the pulse has passed through it, an elecromagnetic-polarization
(polariton) oscillation persists and generally slowly damps out. We demonstrate that non-trivial ENZ features like the enhancement of the longitudinal
electric field component are still observable in the time-domain. We also find that the above mentioned phenomenology is not observed for optical pulses
with carrier frequencies far from the ENZ condition. Thus, in order to grasp the underpinning physical mechanisms responsible for transient trapping in
the ENZ regime, we analytically investigate the scattering features of the Lorentz slab by studying the virtual leaky modes of the structure. We recognize
that a set of polariton modes with reduced transverse group velocity ($v_g\simeq c/100$, where $c$ is the speed of light in vacuum) is excited. Indeed,
the plasma frequency plays the role of a cut-off frequency and polaritons in the ENZ regime are intrinsically characterized by a reduced group velocity.
Thus, we are able to interpret the transient trapping by means of the excitation of slow polariton modes that are damped off due to medium absorption and
radiation leakage in the outer medium.

The paper is organized as follows. In Sec.II we report the results of numerical finite difference time domain (FDTD) simulations, comparing the distinct
phenomenologies occuring in ENZ and standard dielectric regimes. In Sec.III we analytical investigate the virtual leaky modes of the structure, we address
their properties and we discuss their role in the interpretation of numerical results are developed in section III. In Sec.IV we draw our conclusions.

\section{Finite Difference Time Domain analysis of the time-domain ENZ regime}

\subsection{Pulse scattering by a Lorentz slab}
Let us consider the scattering interaction sketched in Fig.1(a), where an electromagnetic pulse is launched along the $z$-axis in vacuum
and orthogonally impinges on the surface of a dielectric slab. The pulse is a Transverse Magnetic (TM) excitation, with electric
$E_x(x,z,t)$, $E_z(x,z,t)$ and magnetic $H_y(x,z,t)$ field components. The initial profile profile of the transverse electric
component is
\begin{equation} \label{pulse}
E_x (x,z,t=0) = E_0 e^{-\frac{x^2}{w_x^2}}  e^{-\frac{(z-z_0)^2}{w_z^2}} \sin \left[ \frac{\bar{\omega}}{c} (z-z_0) \right].
\end{equation}
This field is spatially confined both along the $x$- and $z$- axis, $w_x$ and $w_z$ being its transverse and longitudinal widths,
respectively, it is centered at the point $(x,z)=(0,z_0)$ and it is longitudinally modulated with period
$\bar{\lambda} = 2\pi c/ \bar{\omega} $. Hereafter we will focus on very long pulses such that
$w_z \gg c/(\bar{\omega})$. Thus, the quasi-monochromatic condition $\delta \omega / \bar{\omega} \ll 1$ is satisfied, where
$\bar{\omega}$ is the pulse central frequency and $\delta \omega \simeq c/w_z$ is the spectral width.

\begin{figure}
\centering
\begin{center}
\includegraphics[width=0.45\textwidth]{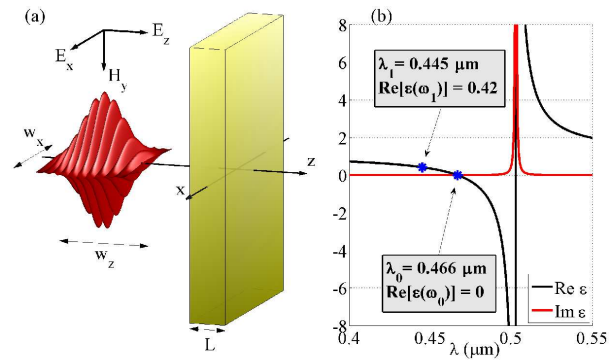}
\caption{(a) Interaction geometry of the pulse colliding onto the dielectric slab. (b) Real and imaginary part of the slab
dielectric permittivity for the Lorentz parameters used in the FDTD analysis as a function of the wavelength.}
\label{Fig1}
\end{center}
\end{figure}

The dielectric slab has width $L$, it is centered at $(x,z)=(0,0)$ and we assume that, in the presence of the external electric field $\bf E$, the
dynamics of its dielectric polarization is governed by the Lorentz oscillator model
\begin{equation} \label{Lorentz}
\frac{d^2{\bf P}}{dt^2} + \gamma \frac{d{\bf P}}{dt} + \omega_e^2 {\bf P} = \epsilon_0 f_e {\bf E} ,
\end{equation}
where $\omega_e$ is the resonant angular frequency, $\gamma$ is the damping constant, $f_e$ is the oscillator strength and $\epsilon_0$ is the dielectric
permittivity of vacuum. It is well known that Eq.(\ref{Lorentz}) leads, in the frequency domain, to the constitutive relation $\tilde{\bf D}_\omega =
\epsilon_0 \epsilon(\omega) \tilde{\bf E}_\omega$ where $\tilde{f}_\omega = \int_{-\infty}^{+\infty} dt e^{i \omega t} f(t)$ is the Fourier transform of
$f(t)$, ${\bf D} = \epsilon_0 {\bf E} + {\bf P}$ is the displacement field vector and $\epsilon = 1 + f_e/(\omega_e^2-i\gamma\omega-\omega^2)$ is the
frequency dependent medium dielectric permittivity. The realistic model of Eq.(\ref{Lorentz}) is particularly accurate for describing the medium
dielectric response to fields with frequencies close to the resonant frequency $\omega_e$ (so that contributions due to other resonances can be
neglected). Therefore, the Lorentz model is particularly suitable for our analysis since we are here concerned with quasi-monochromatic pulses whose
carrier frequency $\bar{\omega}$ coincides with (or is close to) the frequency
\begin{equation} \label{omega0}
\omega_0 = \sqrt{\frac{1}{2} \left[\left(2\omega_e^2-\gamma^2+f_e\right) + \sqrt{f_e^2 + \gamma^2 \left(\gamma^2 - 4 \omega_e^2 - 2 f_e \right)} \right]}
\end{equation}
where the real part of the permittivity vanishes, i.e. ${\rm Re} \left[\epsilon(\omega_0)\right] = 0$, the so called epsilon-near-zero (ENZ) regime.

\begin{figure*}
\includegraphics[width=0.95\textwidth]{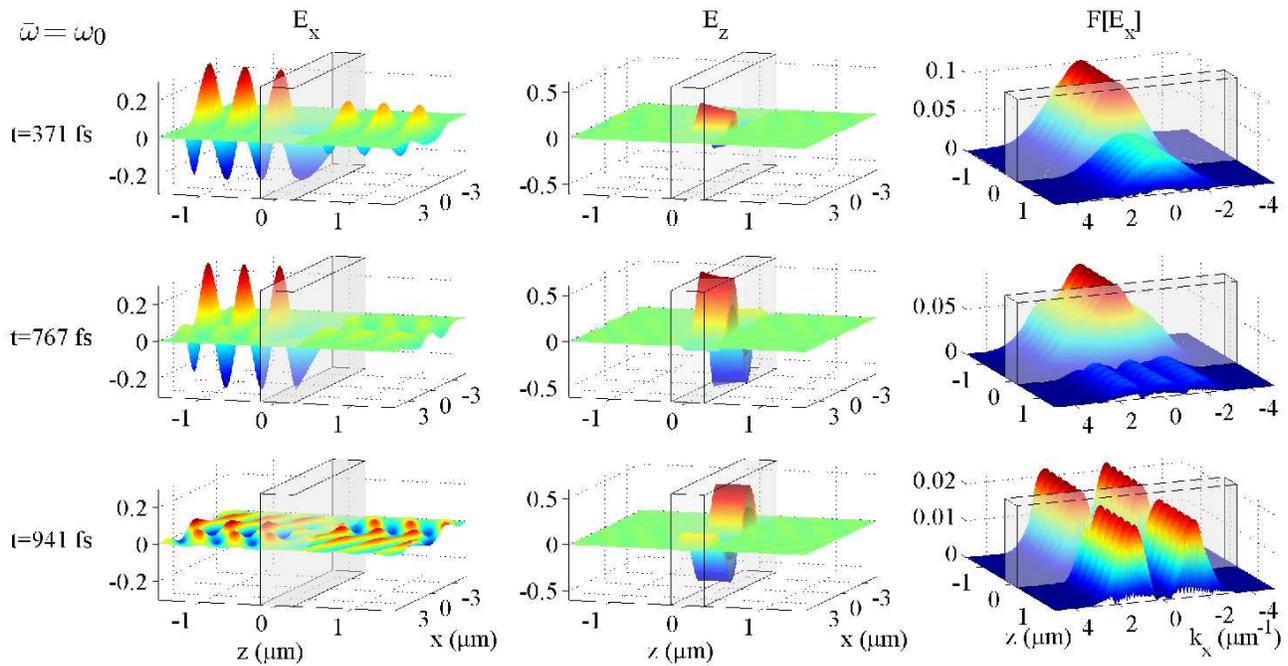}
\caption{Results of FDTD simulation pertaining the interaction of the pulse $0$ (with $\bar{\omega}=\omega_0)$ with the slab. The fields $E_x$, $E_z$ and
the Fourier transform $F[E_x]$ are captured at three different time steps corresponding to the three rows. In the first and the second time steps the
pulse peak is just behind and beyond the slab, respectively, whereas in the third time step the pulse has completely left the slab. Note the large $E_z$
component within the slab (signature of the epsilon near zero regime) and the persisting and pulse-free polariton oscillation in the third time step.}
\end{figure*}

We have performed the numerical analysis of the pulse-slab collision by means of a Finite Difference Time Domain (FDTD) scheme where the polarization
dynamics of Eq.(\ref{Lorentz}) are coupled to Maxwell equations for the TM field. Specifically, in order to isolate the relevant phenomenology
characterizing the ENZ regime we have analyzed through FDTD simulations two different situations where the same dielectric slab is hit by two spatially
equal pulses with different carrier frequencies $\bar{\omega}$: the first (pulse $0$) is such that $\bar{\omega}=\omega_0$ so that it is suitable to scan
the slab behavior in its ENZ regime; the second (pulse $1$) has $\bar{\omega}=\omega_1$ for which ${\rm Re}[\epsilon(\omega_1)] > 0$ so that it
experiences standard dielectric behavior.

In view of the generality and ubiquity of the Lorentz model of Eq.(\ref{Lorentz}), we have chosen for our numerical simulations a realistic medium with
Lorentz parameters $\omega_e = 3.75 \cdot 10^{15} \: Hz$, $\gamma = 1.50 \cdot 10^{12} \: sec^{-1}$ and $f_e = 2.25 \cdot 10^{30} sec^{-2}$ in order to
deal with optical pulses in the visible spectrum, the resonant frequency corresponding to the wavelength $\lambda_e = 2\pi c/\omega_e = 0.502 \: \: \mu
m$. For such parameters Eq.(\ref{omega0}) yields $\omega_0 = 4.03 \cdot 10^{15} \: Hz$ and we have set $\omega_1 = 4.23 \cdot 10^{15} \: Hz$ (for which
${\rm Re}[\epsilon(\omega_1)]=0.42$), the two frequencies corresponding to the wavelengths $\lambda_0 = 2\pi c/\omega_0 = 0.466 \: \: \mu m$ and
$\lambda_1 = 2\pi c/\omega_1 = 0.445 \: \: \mu m$ respectively. In Fig.1(b) we plot the real and imaginary parts of the dielectric permittivity for the
chosen Lorentz parameters as functions of the wavelength $\lambda = 2\pi c /\omega$, indicating the carrier wavelengths $\lambda_0$ and $\lambda_1$ that
characterize the two pulses. We have chosen a slab width $L=0.41 \: \: \mu m$ to minimize the pulse propagation features and to effectively highlight the
impact of the medium polarization on field dynamics. We set $z_0 = - 150 \: \: \mu m$, so that the pulse peak reaches the slab $500 \: fs$ after it has
been launched. Pulse widths are chosen so that $w_x = 1 \: \: \mu m$ and $w_z = 60 \: \: \mu m$: the former is comparable with the central wavelength of
the pulse in order to provide the optical beam with a non negligible longitudinal field component $E_z$ (see below) whereas the latter corresponds to a
temporal width $T=w_z/c = 200 \: \: fs$ and a spectral width $\delta \omega \simeq 1 / T = 5 \cdot 10^{12} \: Hz$ so that the pulses are in the
quasi-monochromatic regime.

\begin{figure*}
\includegraphics[width=0.95\textwidth]{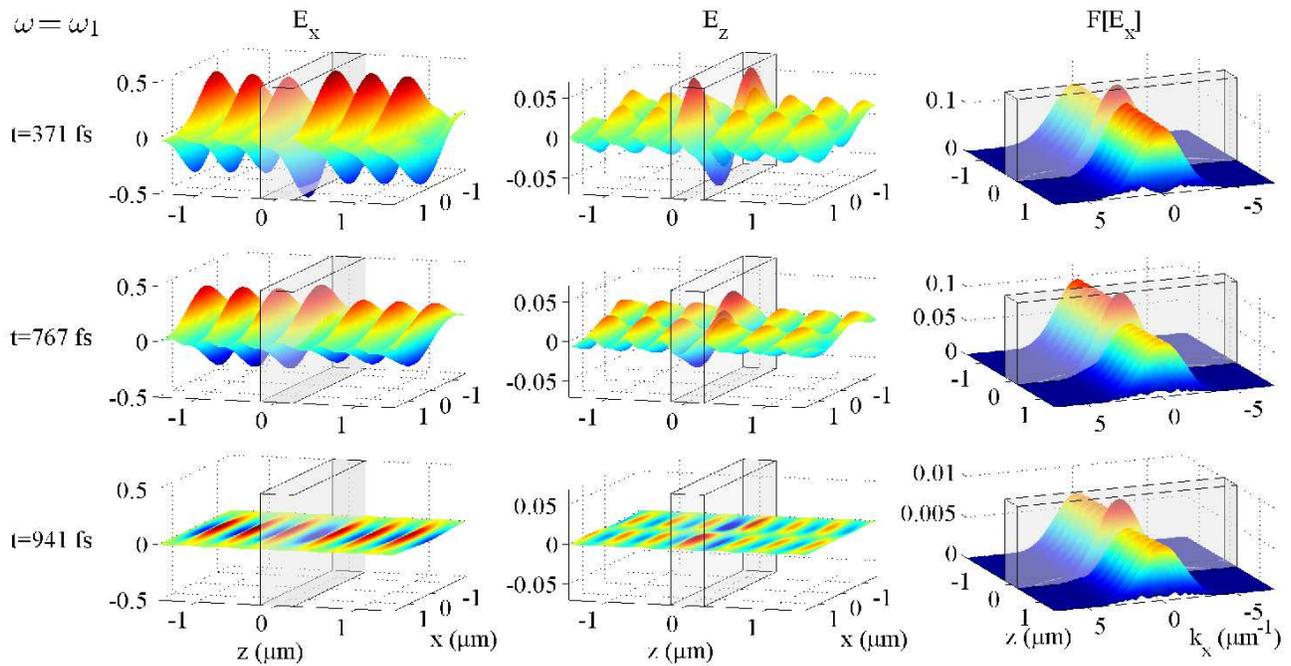}
\caption{Results of FDTD simulation pertaining the interaction of the pulse $1$ (with $\bar{\omega}=\omega_1)$ with the slab.
The fields $E_x$, $E_z$ and the Fourier transform $F[E_x]$, for comparison purposes, are captured at same time steps considered
in Fig.2. Note that both $E_x$ and $F[E_x]$ are bell-shaped, that magnitudes of $E_z$ within the slab and in vacuum are comparable
and that no residual polariton oscillation has been produced by pulse passage.}
\end{figure*}

\subsection{Pulse $0$ scattering}
In Fig.2 we report the main results of the FDTD simulation dealing with the interaction of the pulse $0$ with carrier frequency $\bar{\omega}=\omega_0$
with the Lorentz slab represented in the figure by the semitransparent rectangular blocks. The first two columns of the figure contain the plots of
$E_x(x,z,t)$, $E_z(x,z,t)$ as functions of $(x,z)$ whereas the third contains the Fourier transform $F[E_x](k_x,z,t) = \int_{-\infty}^{+\infty} dx
e^{ik_xx} E_x(x,z,t)$ as a function of $(k_x,z)$; each row of the figure corresponds to a selected simulation time step. At the first time step (first row
of Fig.(2)), $t=371 \: fs$, the pulse is fully interacting with the slab (the pulse peak being about to hit the slab at $t= 500 \: fs$) and the
electromagnetic field is characterized by standard reflection/transmission features; in particular both the reflected and transmitted pulses have a
transverse bell-shaped spatial profile and accordingly the Fourier transform $F[E_x]$ is peaked as well. Note however that, even in this early transient
stage of the interaction, within the slab the longitudinal component $E_z$ is comparable with $E_x$ and much greater than its vacuum counterpart. Such an
enhancement of the electric field component perpendicular to the slab is a feature typically associated with {\it monochromatic} ENZ regime, arising as a
consequence of the continuity of the displacement field component perpendicular to the interface \cite{Vincenti_PRA_2011,Ciattoni_PRA_2012}. Therefore
this is the first evidence that the ENZ regime can effectively be observed in thoroughly realistic Lorentz slabs by means of an equally realistic
scattering interaction configuration. The second row of Fig.2 corresponds to the time step $t=767 \: fs$ a time when the incoming pulse (if freely
propagating) would have passed behind the slab (its temporal width being $200 \: fs$). Note that the longitudinal component $E_z$ is even greater than the
previous time step, testimony to the fact that the ENZ regime also occurs in the time-domain. The transverse component $E_x$ shows novel spatial features,
even more evidently displayed by its Fourier transform $F[E_x]$ which is no longer bell-shaped and characterized by a complex multi-structured profile.
The third row of Fig.2 considers a later time step $t=941 \: fs$ much longer than the time spent by the pulse to fully travel into the slab and leave it.
At this time step, the longitudinal component $E_z$ is {\it still} very large within the slab and the transverse component $E_x$ displays novel and
unexpected features: it is symmetric under the reflection $z \rightarrow -z$, it is not transversally bell-shaped and its Fourier transform $F[E_x]$
displays two peaks at the sides of $k_x=0$. Such phenomenology can be interpreted only by assuming that the interaction of pulse $0$ with the slab is
accompanied by the excitation of a polariton mode whose oscillation lasts a time much longer than the pulse-slab interaction time.

\begin{figure*}
\includegraphics[width=0.47\textwidth]{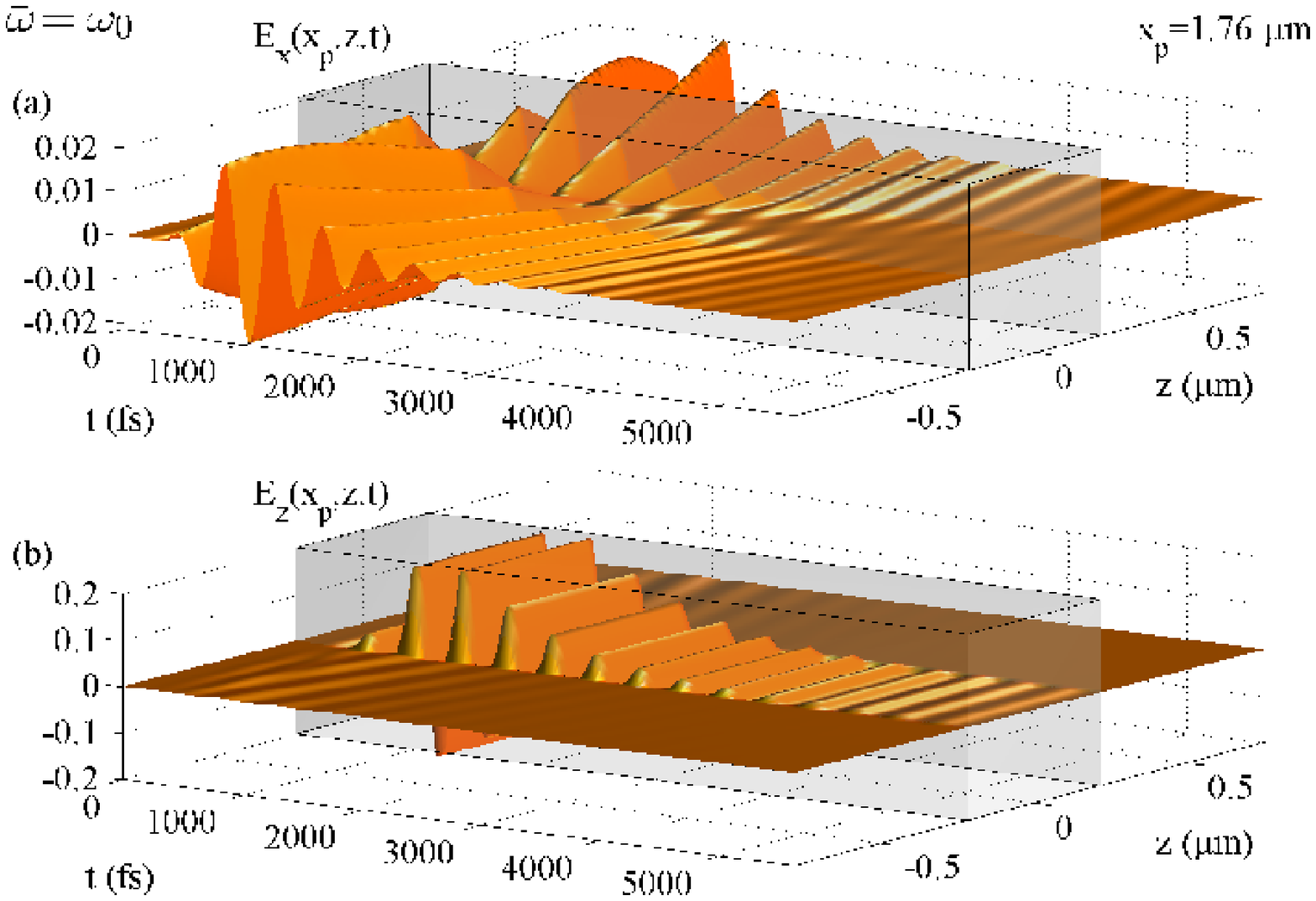}
\includegraphics[width=0.47\textwidth]{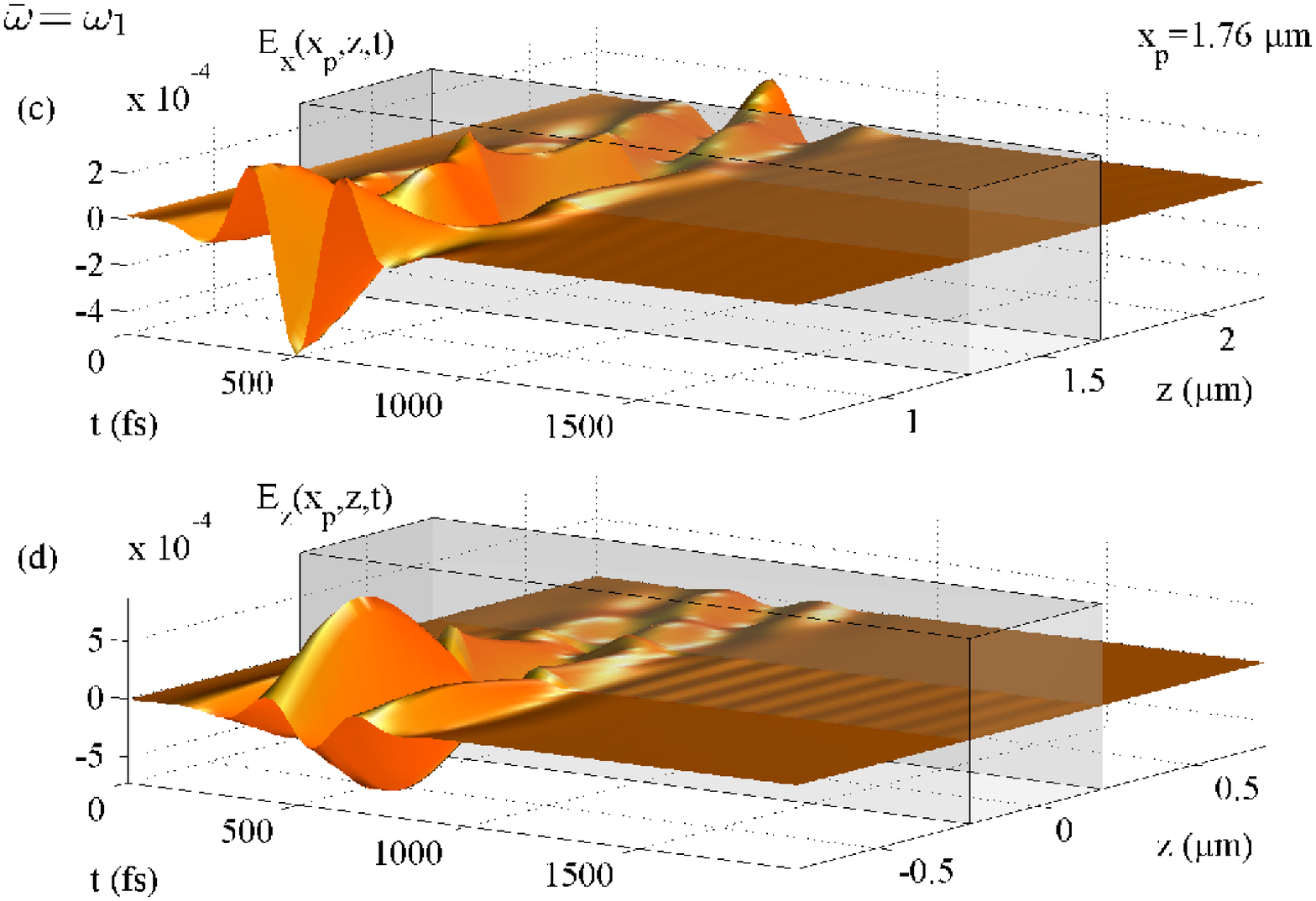}
\caption{FDTD predictions about the fields $E_x$ and $E_z$, for both pulse $0$ and pulse $1$ as function of $(z,t)$ at a fixed plane
$x=x_p=1.76 \: \mu$. Pulse $0$ triggers a novel mechanism of metastable light trapping since its passage produces a strong and damped
polariton oscillation which is absent in the case of pulse $1$.}
\end{figure*}

\subsection{Pulse $1$ scattering}
In order to appreciate the novelty of the above discussed time-domain ENZ phenomenology, we now discuss the interaction of pulse $1$ with the slab, its
carrier frequency being associated to {\it standard} slab dielectric behavior. In Fig.3 we report the results of the FDTD simulation relative to pulse $1$
and, for comparison purposes, we have given Fig.3 the same structure as Fig.2 with the same fields at same time steps. Remarkably, both $E_x$ and $F[E_x]$
are everywhere and always bell-shaped, while the magnitude of $E_z$ within the slab is comparable with its vacuum magnitude. At the last time step the
slab hosts no residual polariton oscillation resulting from the pulse passage. This is precisely the standard expected phenomenology of the reflection and
transmission of the pulse by a dielectric slab, and the comparison with the results of Fig.2 proves that the phenomenology it contains is a manifestation
of the time-domain ENZ regime

\subsection{Transient trapping in the time-domain epsilon near zero regime}
In addition to the remarkable fact that the same features of the monochromatic ENZ regime characterize its time domain counterpart (e.g. the slab hosts a
pronounced enhancement of the field $E_z$), the results discussed in the previous sections also clearly reveal that the scattering situation leads to the
unique excitation of a polariton mode. In order to show more explicitly such a phenomenology, in Fig.4 we have plotted the fields $E_x$ and $E_z$, for
both pulse $0$ and pulse $1$ as functions of $(z,t)$ at a fixed plane $x=x_p=1.76 \: \mu m$. The evident feature that emerges is that pulse $0$ (see
subplots (a) and (b) of Fig.4) produces a strong and damped electromagnetic self-oscillation persisting a time (about $5500 \: fs$) much longer than the
probing pulse duration ($200 \: fs$), self-oscillation which is conversely not produced by pulse $1$ (see subplots (c) and (d) of Fig.4), whose
electromagnetic track fades within the slab just after it has left the medium (at about $t=1000 \: fs$). We conclude that, in the ENZ regime, the pulse
travelling through the slab triggers a novel mechanism of transient light trapping.

\section{Theoretical analysis of time-domain ENZ regime}

\subsection{Polariton virtual modes analysis}
From the above discussed phenomenology, it is evident that a quasi-monochromatic pulse with a spectrum centered at the zero of the real part of the slab
permittivity excites a polariton mode that lasts a time much longer than the pulse-slab interaction time. In order to rigorously prove this statement and
gain deeper understanding of the underpinning physical mechanisms that support the time-domain ENZ regime, in this section we analyze the exact
quasi-steady modes (virtual modes) of the slab. In our analysis we fully take into account damping processes, which include medium absorbtion and
radiation leakage in vacuum, adopting the complex frequency approach \cite{Archambault_PRB_2009}. We start our analysis from the curl Maxwell equations
for TM fields
\begin{eqnarray} \label{Maxwell}
-\frac{\partial E_z}{\partial x}+ \frac{\partial E_x}{\partial z} &=& -\mu_0 \frac{\partial H_y}{\partial t}, \nonumber \\
-\frac{\partial H_y}{\partial z} & = & \epsilon_0 \frac{\partial E_x}{\partial t} +\frac{\partial P_x}{\partial t},  \\
\frac{\partial H_y}{\partial x} & = & \epsilon_0 \frac{\partial E_z}{\partial t} +\frac{\partial P_z}{\partial t}, \nonumber
\end{eqnarray}
where the polarization ${\bf P}(x,z,t) = P_x(x,z,t) \hat{\bf e}_x + P_z(x,z,t) \hat{\bf e}_z$ satisfies Eq.(\ref{Lorentz}) within the slab ($|z|<L/2$) and
it vanishes outside the slab ($|z|>L/2$). We take the Ansatz $A_j (x,z,t) = {\rm Re} \left[a_j(z) e^{i \left(k_x x - \Omega t \right)} \right]$ for every
field component ($A=E,H,P$, $a=e,h,p$ and $j=x,z$), where $k_x$ is the (real) transverse wavevector and $\Omega = \omega - i \Gamma$ is the complex
angular frequency with $\Gamma >0$ so that only damping modes are considered. Owing to the mutual temporal evolution of the electromagnetic field $({\bf
E},{\bf H})$ and of the polarization field ${\bf P}$, the Ansatz effectively amounts to considering polariton virtual modes. The magnetic field can be
expressed in terms of the electric field components $h_y = -( k_x e_z + i \partial_z e_x )/\left( \mu_0 \Omega \right)$ so that Maxwell's equations reduce
to
\begin{eqnarray}
e_z = \frac{ i k_x }{\displaystyle \frac{\Omega^2}{c^2} \tilde{\epsilon}(\Omega,z) - k_x^2} \frac{d e_x}{d z} ,  \label{TmEqEz} \\
\frac{d^2 e_x}{d z^2} + \left[  \frac{\Omega^2}{c^2} \tilde{\epsilon}(\Omega,z) -k_x^2 \right] e_x = 0 , \label{TmEqEx}
\end{eqnarray}
where $\tilde{\epsilon}(\Omega,z) = \epsilon(\Omega)\theta(L/2-|z|) + \theta(|z|-L/2)$ ($\theta(z)$ being the Heaviside step function) is the
$z$-dependent dielectric profile. It is worth stressing that the permittivity $\epsilon$ is evaluated at the complex frequency $\Omega$. The general
solution of Eqs.(\ref{TmEqEz},\ref{TmEqEx}) is explicitly given by
\begin{eqnarray} \label{modes}
e_x (z) & = & C \left\{ \begin{array}{lr}
  \displaystyle \Theta  e^{ -i \Xi K \left( z + \frac{L}{2} \right)} & z < - \frac{L}{2}, \\
  \displaystyle \frac{\displaystyle  e^{ikz} + \Theta e^{-ikz} }{\displaystyle e^{ik\frac{L}{2}} + \Theta e^{-ik\frac{L}{2}}} & -\frac{L}{2} \le z \le \frac{L}{2}, \\
  \displaystyle e^{ i \Xi K \left(z - L/2 \right)} & z > \frac{L}{2}, \end{array} \right.  \nonumber \\ && \\
e_z (z) & = & C \left\{ \begin{array}{cc}
  \displaystyle \Theta \Xi \frac{k_x}{K} e^{-i \Xi K \left(z + \frac{L}{2} \right)}  &  z < - \frac{L}{2}, \\
  \displaystyle \frac{k_x}{k}  \frac{\displaystyle  -  e^{ik z} + \Theta e^{-ikz}}{\displaystyle e^{ik\frac{L}{2}} + \Theta e^{-ik\frac{L}{2}}} & -\frac{L}{2} \le z \le \frac{L}{2},  \\
  \displaystyle - \Xi \frac{k_x}{K}  e^{i \Xi K \left(z-\frac{L}{2}\right)} & z > \frac{L}{2}, \end{array} \right.  \nonumber
\end{eqnarray}
where $K = \sqrt{ \frac{\Omega^2}{c^2} -k_x^2}$, $k = \sqrt{ \frac{\Omega^2}{c^2} \epsilon(\Omega) - k_x^2}$, $C$ is the arbitrary mode amplitude, $\Theta
= \pm 1$ is a parameter that distinguishes the symmetry of the solutions and $\Xi = \pm 1$ is another parameter selecting the sign of the exponentials in
vacuum. By construction, the modal fields in Eqs.(\ref{modes}) already satisfy the continuity of the field component parallel to the slab surface ($e_x$)
at the interfaces $x=\pm L/2$. The boundary conditions (BCs) for the continuity of the displacement field component ($\tilde{\epsilon} e_z$)perpendicular
to the interfaces $x=\pm L/2$ yield  the dispersion relation
\begin{equation} \label{DispRel}
\left( K \epsilon - \Xi k \right) e^{ikL} = \Theta ( K \epsilon + \Xi k),
\end{equation}
which provides the complex frequency $\Omega$ for every given slab thickness $L$ and transverse wave vector $k_x$. We have solved Eq.(\ref{DispRel})
numerically and obtained the allowed $\Omega$ corresponding to different values of $k_x$ using the same slab thickness and Lorentz dispersive parameters
of the slab considered in Section II. In Fig.5 we plot the results for the case $\Theta = -1$ in the complex plane $\Omega$ parametrized through the
wavelength $\lambda$ and the damping constant $\Gamma$ (i.e. $\Omega = 2\pi c / \lambda - i \Gamma$), using circles and stars for the $\Xi = -1$ and $\Xi
= 1$ modes, respectively, and using the marker color to label the value of the corresponding $k_x$. Note that the $\Xi = -1$ and $\Xi = 1$ modes belongs
to two different branches which are characterized by the fact that the $\Xi = -1$ modes have damping constant greater than the $\Xi = +1$ modes. This
property can be easily understood by considering the $z$-component of the non-oscillatory part of the Poynting vector ${\bf S}= {\bf E} \times {\bf H}$
for $z>L/2$:
\begin{equation} \label{Poyting}
\langle S_z \rangle = \Xi e^{-2 \left[ \Xi {\rm Im}(K) \left(z-\frac{L}{2}\right) + \Gamma t \right]} \frac{1}{2}\epsilon_0 {\rm Re}\left(
\frac{\Omega}{K}\right) |C|^2 .
\end{equation}
For $\Xi = -1$, the energy outflows from the slab and the damping of the virtual mode is more rapid since it loses energy through both medium absorption
and radiation leakage. Conversely, for $\Xi = 1$ the electromagnetic energy is dragged into the slab, thus partially compensating for the medium
absorption and consequently decreasing the virtual mode damping time (note that there is also a point where $\Gamma =0$ on the $\Xi=1$ branch
corresponding to the exact balance between medium absorption and radiation drag).
\begin{figure}
\includegraphics[width=0.48\textwidth]{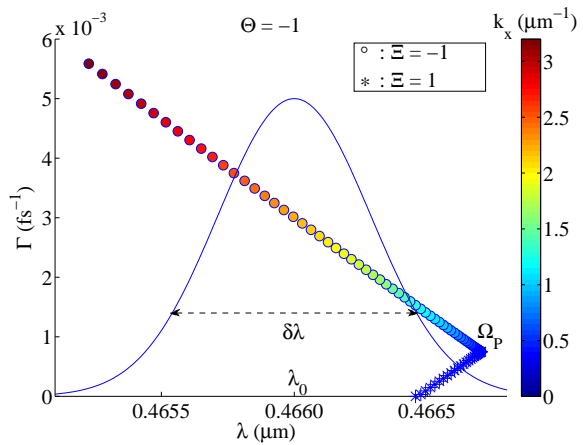}
\caption{Virtual modes of the slab considered in Sec.II for the symmetry $\Theta = -1$ in the complex plane
$\Omega = 2\pi c / \lambda - i \Gamma$. Circles and stars label the $\Xi = -1$ and $\Xi = 1$ modes whereas the marker color labels the
corresponding $k_x$ value. The complex frequency $\Omega_P$ is such that $\epsilon(\Omega_P)=0$. The thin continuous line is the temporal
spectrum of the incoming pulse reported in arbitrary units.}
\end{figure}
It is remarkable that the $\Xi = -1$ and $\Xi = 1$ branches intersect each other at point $\Omega_P$ for $k_x \rightarrow 0$. In this limit the dispersion
relation of Eq.(\ref{DispRel}) (for $\Theta = -1$) reduces to
\begin{equation}
e^{i\frac{\Omega_P}{c}\sqrt{\epsilon(\Omega_P)}} = \frac{\Xi -\sqrt{\epsilon(\Omega_P)} }{\Xi +\sqrt{\epsilon(\Omega_P)} },
\end{equation}
and is satisfied only if $\epsilon(\Omega_P) = 0$. Starting from the Lorentz model, it is straightforward to prove that the permittivity vanishes at
$\Omega_P = \sqrt{f_e + \omega_e^2 - \gamma^2/4}-i \gamma/2$ that, for the above used Lorentz dispersive parameters, yields $\lambda_P = 2\pi c/{\rm
Re}(\Omega_P) = 0.4667 \: \mu m$ and $\Gamma_P = -{\rm Im}(\Omega_P)= 8.3 \cdot 10^{-3} \: fs^{-1}$, precisely matching the point of Fig.5 where the two
branches $\Xi = \pm 1$ intersect each other. In turn, the {\it plasmonic} $\Omega_P$ at which the permittivity vanishes plays a central role in the
analysis of the virtual modes. For the complex frequencies $\Omega$ reported in Fig.5, $|\epsilon(\Omega)| < 0.06$ and therefore, all the obtained modes
with symmetry $\Theta = -1$ imply the time-domain ENZ regime.

In the same portion of the complex plane $\Omega$ we have numerically found no allowed modes for the symmetry $\Theta = 1$. This can be grasped by
expanding both sides of Eq.(\ref{DispRel}) in Taylor series of $\epsilon$ (since $|\epsilon(\Omega)| \ll 1$); at the zeroth order we readily obtain
$e^{-k_xL} = -\Theta$ which is not consistent if $\Theta = 1$ (and which, on the other hand, yields $k_x=0$ for $\Theta = -1$).

\subsection{Interpretation of FDTD results in terms of virtual polariton modes and slow-light regime}
Usually, within the standard real frequency approach, the solutions for the slab modes with $\Xi = - 1$ are disregarded since they are considered
unphysical. Indeed, if $|k_x|> \omega/c$, the solutions with $\Xi = + 1$ represent confined modes propagating along the $x$-direction, while solutions
with $\Xi = - 1$ are unbound modes that diverge at $z\rightarrow\pm\infty$. In addition, the introduction of the complex frequency introduces an inherent
field singularity in the far past $t\rightarrow-\infty$. Due to such intrinsic singularities, it is strictly impossible to rigorously excite a single
virtual mode, its global existence on the whole space-time being unphysical. However, both singularities occur asymptotically and therefore virtual modes
provide a very adequate description of the transient ENZ slab behavior occurring within a spatially bounded region and through a finite time lapse. In
order to prove this statement and to basically provide a theoretical analytical description of the transient light trapping discussed in Sec.2D, in Fig.5
we have superimposed the temporal spectrum profile of the incoming pulse of Eq.(\ref{pulse}) (using the thin continuous line) on the complex-plane virtual
modal structure. Note that, due to its wavelength bandwidth $\delta \lambda = \left( 2\pi c /\omega_0^2 \right) \delta \omega = 5.8 \cdot 10^{-4} \: \mu
m$, the pulse spectrum centered at $\lambda_0$ overlaps a limited portion of the considered complex frequency plane so that, specifically, the sole
virtual modes with $|\lambda - \lambda_0| < \delta \lambda/2$ are actually excited by the considered pulse 0. From Fig.5 it is evident that the excited
virtual modes are characterized by the transverse wavevector $k_x$ spanning the range $1.1 \: \: \mu m^{-1} < k_x < 2.7 \: \: \mu m^{-1}$ and that, due to
the finite bandwidth of the impinging pulse, the excited virtual modes with largest amplitude are those around the central transverse wavevector $k_x =
1.9 \: \mu m$, which corresponds to $\lambda = 2\pi c/{\rm Re}(\Omega)$ close to $\lambda_0 = 0.466$. This observation is in striking agreement with the
results contained in Fig.2, where one can see that the transverse Fourier transform of the field has, at the latest time step, two peaks centered at $k_x
\simeq 1.9  \: \mu m$ and $k_x \simeq -1.9  \: \mu m$ whose width is of the order of $\delta k_x \simeq 0.8 \: \mu m$. Therefore, when the pulse 0
impinges onto the slab, it excites precisely the virtual modes analytically predicted in Sec.IIIA, which are compatible with its spectral structure. As a
further validation of this statement, note that the virtual modes excited by the pulse 0 ($|\lambda - \lambda_0| < \delta \lambda/2$) have a damping
constant $\Gamma$ spanning the range $1.2 \cdot 10^{-3} fs^{-1} < \Gamma < 4.9 \cdot 10^{-3} fs^{-1}$ (see Fig.5), which corresponds to the extinction
time $\tau = 3/\Gamma$ spanning the range $609 fs < \tau < 2604 fs$. Also this prediction based on the above virtual mode analysis is in striking
agreement with the FDTD results since, by looking at panels (a) and (b) of Fig.4, one can see that the electromagnetic excitation persists for a time of
the order of $3000 \: fs$, which is compatible with the maximum extinction time of the excited virtual modes. In addition, the spatial symmetry of the
$\Theta = -1$ virtual polariton modes matches the numerical results displayed in  Fig.4: the transverse field component ($e_x$) is antisymmetric with
respect to the $z=0$ axis, whereas the longitudinal field component ($e_z$) is symmetric.

The final ingredient needed to thoroughly interpret the transient light trapping observed in FDTD simulations is related to the intrinsic slow-light
nature of the phenomenon, which may be preliminarily grasped by considering a bulk Lorentz medium, where transverse plane waves satisfy the dispersion
relation $k(\omega) = (\omega/c)\sqrt{\epsilon(\omega)}$. Neglecting medium absorption ($\gamma \simeq 0$), one finds that the phase velocity is $v_f =
\omega / k$ and the group velocity $v_g=\mathrm{d}\omega/\mathrm{d}k$ is
\begin{equation}
v_g (\omega) = c \left[ \sqrt{\epsilon(\omega)} + \frac{ f_e \omega^2 }{ \sqrt{\epsilon(\omega)} (\omega_e^2-\omega^2)^2 } \right]^{-1} .
\end{equation}
Thus, in the ENZ regime the phase velocity diverges $v_f\rightarrow\infty$ whereas the group velocity tends to zero $v_g\rightarrow 0$. Even though the
subwavelength Lorentz slab used in our FDTD simulations is not a bulk medium and absorption has not been neglected, the rough argument above still
predicts the correct outcome. Indeed, by numerically solving the dispersion relation of Eq.(\ref{DispRel}) without neglecting losses one finds that, at
the optical wavelength $\lambda_0$, the transverse phase velocity of virtual polariton modes is superluminal $v_f = \omega/k_x \simeq 10 c$, while the
transverse group velocity is extremely reduced $v_g = \mathrm{d}\omega/\mathrm{d}k_x \simeq c/100$. For this reason, it is now clear how the virtual
polariton modes, once excited, do not disperse quickly in the $x$-direction and remain quasi-trapped within the slab owing to the tremendously reduced
temporal dynamics. We conclude that the above described transient light trapping can be fully interpreted and physically understood by means of slow
polariton modes supported by the slab.

\subsection{Volume plasmons}

Although the above discussed numerical and analytical analysis of the transient light trapping characterizing the time-domain ENZ regime is quite
exhaustive, we now discuss its connection with the purely longitudinal modes, either {\it volume plasmons} (collective oscillations of electrons) or {\it
volume phonons} (collective oscillations of ions), which the Lorentz medium can support. Hereafter we focus on volume plasmons, considering an unbounded
bulk Lorentz medium where the TM electromagnetic and polarization dynamics are described by Eqs.(\ref{Lorentz},\ref{Maxwell}). For the plane-wave Ansatz
$A_j (x,z,t) = {\rm Re} \left[a_j e^{i \left(k_x x + k_z z - \Omega t \right)} \right]$, where $A=E,H,P$, $a=e,h,p$, $j=x,z$, $k_x,k_z$ are the (real)
wavevector components and $\Omega$ is the generally complex frequency one gets
\begin{eqnarray} \label{amplit}
\mu_0 \Omega h_y = -k_x e_z + k_z e_x, \nonumber \\
k_z h_y = \Omega \epsilon_0 \epsilon(\Omega) e_x,  \\
k_x h_y = -\Omega \epsilon_0 \epsilon(\Omega) e_z, \nonumber
\end{eqnarray}
where $\epsilon(\Omega)$ is the dielectric permittivity with complex frequency $\Omega$. Volume plasmons are purely longitudinal electric oscillations
owing to to the collective motion of electrons and are not accompanied by the generation of magnetic field, a feature that for plane waves amounts to the
collinearity of the wave vector ${\bf k} = k_x \hat{\bf e}_x + k_z \hat{\bf e}_z$ and the electric field ${\bf E} = k_x \hat{\bf E}_x + k_z \hat{\bf
E}_z$. Therefore, imposing the condition ${\bf k} \times {\bf E} =0$, i.e. $-k_x e_z + k_z e_x=0$, Eqs.(\ref{amplit}) readily yield $h_y=0$ and
$\epsilon(\Omega_P) =0$. Thus, volume plasmons are inherently involved in the time-domain ENZ regime we are considering in this paper. However, it is
worth noting that the virtual modes of the Lorentz slab are polaritons, entities fundamentally different from volume plasmons (or volume phonons). Indeed,
a volume plasmon is strictly characterized by the condition $\epsilon(\Omega)=0$ that implies the severe dispersion $\Omega=\Omega_P$ and, in the presence
of the slab boundaries at $z=\pm L/2$, inevitably leads to the inconsistency $E_z \rightarrow \infty$ within the slab unless $E_z = 0$ in the outer
medium. This is consistent with the well-known {\it impossibility} to excite volume plasmons by means of light. On the other hand, from Eq.(\ref{modes})
one can see that the virtual polariton mode component $E_z$ neither vanishes outside the slab nor diverges within it. This is because for polaritons the
dispersion relation of Eq.(\ref{DispRel}) is not as stiff as the volume plasmon dispersion $\Omega = \Omega_P$ and is satisfied also for $\Omega \neq
\Omega_P$. In addition the volume plasmon is a purely electric oscillation with strictly null magnetic field, whereas the considered virtual polariton
modes are accompanied by a magnetic field. In turn, even though TM polariton modes and volume plasmons occur in the same spectral region and are
accidentally connected by the fact that in the limit $L>>\lambda$ the Lorentz slab is almost equivalent to a bulk medium, conceptually they are very
distinct entities. In view of this, we conclude remarking that volume plasmons can not be excited by classical light and that the absorption peak observed
in experiments \cite{Hoeflich_PRL_2009,Hoeflich_PRL_2010} is due to the excitation of virtual polariton modes, confirming the results given in Refs.
\cite{Henrard_SynthMet_1999,Henrard_PRL_2010}.

\section{Conclusions}

In conclusion we have investigated both numerically and analytically the properties of the time-domain ENZ regime. Specifically we have considered a
dielectric slab whose polarization dynamics has been described through the realistic and ubiquitous Lorentz model and we have analyzed its interaction
with quasi-monochromatic and spatially confined pulses with carrier frequencies close to the crossing point of the permittivity real part. The FDTD
analysis has shown that the pulse is able to excite a polarization-electromagnetic (polariton) oscillation which is damped and persists for a time
generally longer than the effective time required by the pulse for passing through the slab. The underlying nature of this excitation has been elucidated
through the analysis of the slab virtual modes that turn out to be located in a portion of the complex frequency plane close to the plasmonic frequency
characterizing plasmon/phonons longitudinal volume excitations. Remarkably, due to this spectral property, both the group velocity and the transverse
velocity (parallel to the slab) of each virtual mode turn out to be very small and therefore, the time-domain ENZ regime be naturally regarded as a novel
platform for discussing and investigating a plethora of slow-light phenomena.


\begin{thebibliography}{35}
\expandafter\ifx\csname natexlab\endcsname\relax\def\natexlab#1{#1}\fi \expandafter\ifx\csname bibnamefont\endcsname\relax
  \def\bibnamefont#1{#1}\fi
\expandafter\ifx\csname bibfnamefont\endcsname\relax
  \def\bibfnamefont#1{#1}\fi
\expandafter\ifx\csname citenamefont\endcsname\relax
  \def\citenamefont#1{#1}\fi
\expandafter\ifx\csname url\endcsname\relax
  \def\url#1{\texttt{#1}}\fi
\expandafter\ifx\csname urlprefix\endcsname\relax\def\urlprefix{URL }\fi \providecommand{\bibinfo}[2]{#2} \providecommand{\eprint}[2][]{\url{#2}}

\bibitem[{\citenamefont{Tsakmakidis et~al.}(2007)\citenamefont{Tsakmakidis,
  Boardman, and Hess}}]{Tsakmakidis_NatLett_2007}
\bibinfo{author}{\bibfnamefont{K.~L.} \bibnamefont{Tsakmakidis}},
  \bibinfo{author}{\bibfnamefont{A.~D.} \bibnamefont{Boardman}},
  \bibnamefont{and} \bibinfo{author}{\bibfnamefont{O.}~\bibnamefont{Hess}},
  \bibinfo{journal}{Nature Letters} \textbf{\bibinfo{volume}{450}},
  \bibinfo{pages}{397} (\bibinfo{year}{2007}).

\bibitem[{\citenamefont{Boyd}(2009)}]{Boyd_JModPhys_2009}
\bibinfo{author}{\bibfnamefont{R.~W.} \bibnamefont{Boyd}},
  \bibinfo{journal}{Journal of Modern Optics} \textbf{\bibinfo{volume}{56}},
  \bibinfo{pages}{1908} (\bibinfo{year}{2009}).

\bibitem[{\citenamefont{Kim et~al.}(2012)\citenamefont{Kim, Husakou, and
  Herrmann}}]{Kim_OE_2012}
\bibinfo{author}{\bibfnamefont{K.~H.} \bibnamefont{Kim}},
  \bibinfo{author}{\bibfnamefont{A.}~\bibnamefont{Husakou}}, \bibnamefont{and}
  \bibinfo{author}{\bibfnamefont{J.}~\bibnamefont{Herrmann}},
  \bibinfo{journal}{Optics Express} \textbf{\bibinfo{volume}{20}},
  \bibinfo{pages}{25790} (\bibinfo{year}{2012}).

\bibitem[{\citenamefont{Vlasov et~al.}(2005)\citenamefont{Vlasov, O'Boyle,
  Hamann, and McNab}}]{Vlasov_Nature_2005}
\bibinfo{author}{\bibfnamefont{Y.~A.} \bibnamefont{Vlasov}},
  \bibinfo{author}{\bibfnamefont{M.}~\bibnamefont{O'Boyle}},
  \bibinfo{author}{\bibfnamefont{H.~F.} \bibnamefont{Hamann}},
  \bibnamefont{and} \bibinfo{author}{\bibfnamefont{S.~J.} \bibnamefont{McNab}},
  \bibinfo{journal}{Nature} \textbf{\bibinfo{volume}{438}}, \bibinfo{pages}{65}
  (\bibinfo{year}{2005}).

\bibitem[{\citenamefont{Mingaleev et~al.}(2006)\citenamefont{Mingaleev,
  Miroshnichenko, Kivshar, and Busch}}]{Mingaleev_PRE_2006}
\bibinfo{author}{\bibfnamefont{S.~F.} \bibnamefont{Mingaleev}},
  \bibinfo{author}{\bibfnamefont{A.~E.} \bibnamefont{Miroshnichenko}},
  \bibinfo{author}{\bibfnamefont{Y.~S.} \bibnamefont{Kivshar}},
  \bibnamefont{and} \bibinfo{author}{\bibfnamefont{K.}~\bibnamefont{Busch}},
  \bibinfo{journal}{Physical Review E} \textbf{\bibinfo{volume}{74}},
  \bibinfo{pages}{046603} (\bibinfo{year}{2006}).

\bibitem[{\citenamefont{Bajcsy et~al.}(2009)\citenamefont{Bajcsy, Hofferberth,
  Balic, Peyronel, Hafezi, Zibrov, Vuletic, and Lukin}}]{Bajcsy_PRL_2009}
\bibinfo{author}{\bibfnamefont{M.}~\bibnamefont{Bajcsy}},
  \bibinfo{author}{\bibfnamefont{S.}~\bibnamefont{Hofferberth}},
  \bibinfo{author}{\bibfnamefont{V.}~\bibnamefont{Balic}},
  \bibinfo{author}{\bibfnamefont{T.}~\bibnamefont{Peyronel}},
  \bibinfo{author}{\bibfnamefont{M.}~\bibnamefont{Hafezi}},
  \bibinfo{author}{\bibfnamefont{A.~S.} \bibnamefont{Zibrov}},
  \bibinfo{author}{\bibfnamefont{V.}~\bibnamefont{Vuletic}}, \bibnamefont{and}
  \bibinfo{author}{\bibfnamefont{M.~D.} \bibnamefont{Lukin}},
  \bibinfo{journal}{Physical Review Letters} \textbf{\bibinfo{volume}{102}},
  \bibinfo{pages}{203902} (\bibinfo{year}{2009}).

\bibitem[{\citenamefont{Afshar and Monro}(2009)}]{Afshar_OE_2009}
\bibinfo{author}{\bibfnamefont{S.~V.} \bibnamefont{Afshar}} \bibnamefont{and}
  \bibinfo{author}{\bibfnamefont{T.~M.} \bibnamefont{Monro}},
  \bibinfo{journal}{Optics Express} \textbf{\bibinfo{volume}{17}},
  \bibinfo{pages}{2298} (\bibinfo{year}{2009}).

\bibitem[{\citenamefont{Marini et~al.}(2011)\citenamefont{Marini, Hartley,
  Gorbach, and Skryabin}}]{Marini_PRA_2011}
\bibinfo{author}{\bibfnamefont{A.}~\bibnamefont{Marini}},
  \bibinfo{author}{\bibfnamefont{R.}~\bibnamefont{Hartley}},
  \bibinfo{author}{\bibfnamefont{A.~V.} \bibnamefont{Gorbach}},
  \bibnamefont{and} \bibinfo{author}{\bibfnamefont{D.~V.}
  \bibnamefont{Skryabin}}, \bibinfo{journal}{Physical Review A}
  \textbf{\bibinfo{volume}{84}}, \bibinfo{pages}{063839}
  (\bibinfo{year}{2011}).

\bibitem[{\citenamefont{Koos et~al.}(2007)\citenamefont{Koos, Jacome, Poulton,
  Leuthold, and Freude}}]{Koos_OE_2007}
\bibinfo{author}{\bibfnamefont{C.}~\bibnamefont{Koos}},
  \bibinfo{author}{\bibfnamefont{L.}~\bibnamefont{Jacome}},
  \bibinfo{author}{\bibfnamefont{C.}~\bibnamefont{Poulton}},
  \bibinfo{author}{\bibfnamefont{J.}~\bibnamefont{Leuthold}}, \bibnamefont{and}
  \bibinfo{author}{\bibfnamefont{W.}~\bibnamefont{Freude}},
  \bibinfo{journal}{Optics Express} \textbf{\bibinfo{volume}{15}},
  \bibinfo{pages}{5976} (\bibinfo{year}{2007}).

\bibitem[{\citenamefont{Palomba and Novotny}(2008)}]{Palomba_PRL_2008}
\bibinfo{author}{\bibfnamefont{S.}~\bibnamefont{Palomba}} \bibnamefont{and}
  \bibinfo{author}{\bibfnamefont{L.}~\bibnamefont{Novotny}},
  \bibinfo{journal}{Physical Review Letters} \textbf{\bibinfo{volume}{101}},
  \bibinfo{pages}{056802} (\bibinfo{year}{2008}).

\bibitem[{\citenamefont{Kauranen and Zayats}(2012)}]{Kauranen_NatPhot_2012}
\bibinfo{author}{\bibfnamefont{M.}~\bibnamefont{Kauranen}} \bibnamefont{and}
  \bibinfo{author}{\bibfnamefont{A.~V.} \bibnamefont{Zayats}},
  \bibinfo{journal}{Nature Photonics} \textbf{\bibinfo{volume}{6}},
  \bibinfo{pages}{737} (\bibinfo{year}{2012}).

\bibitem[{\citenamefont{Ciattoni et~al.}(2010)\citenamefont{Ciattoni, Rizza,
  and Palange}}]{Ciattoni_PRA_2010}
\bibinfo{author}{\bibfnamefont{A.}~\bibnamefont{Ciattoni}},
  \bibinfo{author}{\bibfnamefont{C.}~\bibnamefont{Rizza}}, \bibnamefont{and}
  \bibinfo{author}{\bibfnamefont{E.}~\bibnamefont{Palange}},
  \bibinfo{journal}{Physical Review A} \textbf{\bibinfo{volume}{81}},
  \bibinfo{pages}{043839} (\bibinfo{year}{2010}).

\bibitem[{\citenamefont{Ciattoni et~al.}(2011)\citenamefont{Ciattoni, Rizza,
  and Palange}}]{Ciattoni_PRA_2011}
\bibinfo{author}{\bibfnamefont{A.}~\bibnamefont{Ciattoni}},
  \bibinfo{author}{\bibfnamefont{C.}~\bibnamefont{Rizza}}, \bibnamefont{and}
  \bibinfo{author}{\bibfnamefont{E.}~\bibnamefont{Palange}},
  \bibinfo{journal}{Physical Review A} \textbf{\bibinfo{volume}{83}},
  \bibinfo{pages}{043813} (\bibinfo{year}{2011}).

\bibitem[{\citenamefont{Vincenti et~al.}(2011)\citenamefont{Vincenti, DeCeglia,
  Ciattoni, and Scalora}}]{Vincenti_PRA_2011}
\bibinfo{author}{\bibfnamefont{M.~A.} \bibnamefont{Vincenti}},
  \bibinfo{author}{\bibfnamefont{D.}~\bibnamefont{DeCeglia}},
  \bibinfo{author}{\bibfnamefont{A.}~\bibnamefont{Ciattoni}}, \bibnamefont{and}
  \bibinfo{author}{\bibfnamefont{M.}~\bibnamefont{Scalora}},
  \bibinfo{journal}{Physical Review A} \textbf{\bibinfo{volume}{84}},
  \bibinfo{pages}{063826} (\bibinfo{year}{2011}).

\bibitem[{\citenamefont{Ciattoni and Spinozzi}(2012)}]{Ciattoni_PRA_2012}
\bibinfo{author}{\bibfnamefont{A.}~\bibnamefont{Ciattoni}} \bibnamefont{and}
  \bibinfo{author}{\bibfnamefont{E.}~\bibnamefont{Spinozzi}},
  \bibinfo{journal}{Physical Review A} \textbf{\bibinfo{volume}{85}},
  \bibinfo{pages}{043806} (\bibinfo{year}{2012}).

\bibitem[{\citenamefont{Powell et~al.}(2009)\citenamefont{Powell, Alu, Edwards,
  Vakil, Kivshar, and Engheta}}]{Powell_PRB_2009}
\bibinfo{author}{\bibfnamefont{D.~A.} \bibnamefont{Powell}},
  \bibinfo{author}{\bibfnamefont{A.}~\bibnamefont{Alu}},
  \bibinfo{author}{\bibfnamefont{B.}~\bibnamefont{Edwards}},
  \bibinfo{author}{\bibfnamefont{A.}~\bibnamefont{Vakil}},
  \bibinfo{author}{\bibfnamefont{Y.~S.} \bibnamefont{Kivshar}},
  \bibnamefont{and} \bibinfo{author}{\bibfnamefont{N.}~\bibnamefont{Engheta}},
  \bibinfo{journal}{Physical Review B} \textbf{\bibinfo{volume}{79}},
  \bibinfo{pages}{245135} (\bibinfo{year}{2009}).

\bibitem[{\citenamefont{Argyropoulos et~al.}(2012)\citenamefont{Argyropoulos,
  Chen, D'Aguanno, Engheta, and Alu}}]{Argyropoulos_PRBB_2012}
\bibinfo{author}{\bibfnamefont{C.}~\bibnamefont{Argyropoulos}},
  \bibinfo{author}{\bibfnamefont{P.-Y.} \bibnamefont{Chen}},
  \bibinfo{author}{\bibfnamefont{G.}~\bibnamefont{D'Aguanno}},
  \bibinfo{author}{\bibfnamefont{N.}~\bibnamefont{Engheta}}, \bibnamefont{and}
  \bibinfo{author}{\bibfnamefont{A.}~\bibnamefont{Alu}},
  \bibinfo{journal}{Physical Review B} \textbf{\bibinfo{volume}{85}},
  \bibinfo{pages}{045129} (\bibinfo{year}{2012}).

\bibitem[{\citenamefont{Enoch et~al.}(2002)\citenamefont{Enoch, Tayeb,
  Sabouroux, Guerin, and Vincent}}]{Enoch_PRL_2002}
\bibinfo{author}{\bibfnamefont{S.}~\bibnamefont{Enoch}},
  \bibinfo{author}{\bibfnamefont{G.}~\bibnamefont{Tayeb}},
  \bibinfo{author}{\bibfnamefont{P.}~\bibnamefont{Sabouroux}},
  \bibinfo{author}{\bibfnamefont{N.}~\bibnamefont{Guerin}}, \bibnamefont{and}
  \bibinfo{author}{\bibfnamefont{P.}~\bibnamefont{Vincent}},
  \bibinfo{journal}{Physical Review Letters} \textbf{\bibinfo{volume}{89}},
  \bibinfo{pages}{213902} (\bibinfo{year}{2002}).

\bibitem[{\citenamefont{Alu et~al.}(2006)\citenamefont{Alu, Bilotti, Engheta,
  and Vegni}}]{Alu_IEEE_2006}
\bibinfo{author}{\bibfnamefont{A.}~\bibnamefont{Alu}},
  \bibinfo{author}{\bibfnamefont{F.}~\bibnamefont{Bilotti}},
  \bibinfo{author}{\bibfnamefont{N.}~\bibnamefont{Engheta}}, \bibnamefont{and}
  \bibinfo{author}{\bibfnamefont{L.}~\bibnamefont{Vegni}},
  \bibinfo{journal}{IEEE Trans. Antennas Propag.}
  \textbf{\bibinfo{volume}{54}}, \bibinfo{pages}{1632} (\bibinfo{year}{2006}).

\bibitem[{\citenamefont{Alu and Engheta}(2005)}]{Alu_PRE_2005}
\bibinfo{author}{\bibfnamefont{A.}~\bibnamefont{Alu}} \bibnamefont{and}
  \bibinfo{author}{\bibfnamefont{N.}~\bibnamefont{Engheta}},
  \bibinfo{journal}{Physical Review E} \textbf{\bibinfo{volume}{72}},
  \bibinfo{pages}{016623} (\bibinfo{year}{2005}).

\bibitem[{\citenamefont{Silveirinha and Engheta}(2006)}]{Silveirinha_PRL_2006}
\bibinfo{author}{\bibfnamefont{M.}~\bibnamefont{Silveirinha}} \bibnamefont{and}
  \bibinfo{author}{\bibfnamefont{N.}~\bibnamefont{Engheta}},
  \bibinfo{journal}{Physical Review Letters} \textbf{\bibinfo{volume}{97}},
  \bibinfo{pages}{157403} (\bibinfo{year}{2006}).

\bibitem[{\citenamefont{Alu et~al.}(2007)\citenamefont{Alu, Silveirinha,
  Salandrino, and Engheta}}]{Alu_PRB_2007}
\bibinfo{author}{\bibfnamefont{A.}~\bibnamefont{Alu}},
  \bibinfo{author}{\bibfnamefont{M.~G.} \bibnamefont{Silveirinha}},
  \bibinfo{author}{\bibfnamefont{A.}~\bibnamefont{Salandrino}},
  \bibnamefont{and} \bibinfo{author}{\bibfnamefont{N.}~\bibnamefont{Engheta}},
  \bibinfo{journal}{Physical Review B} \textbf{\bibinfo{volume}{75}},
  \bibinfo{pages}{155410} (\bibinfo{year}{2007}).

\bibitem[{\citenamefont{Castaldi et~al.}(2012)\citenamefont{Castaldi, Savoia,
  Galdi, Alu, and Engheta}}]{Castaldi_PRB_2012}
\bibinfo{author}{\bibfnamefont{G.}~\bibnamefont{Castaldi}},
  \bibinfo{author}{\bibfnamefont{S.}~\bibnamefont{Savoia}},
  \bibinfo{author}{\bibfnamefont{V.}~\bibnamefont{Galdi}},
  \bibinfo{author}{\bibfnamefont{A.}~\bibnamefont{Alu}}, \bibnamefont{and}
  \bibinfo{author}{\bibfnamefont{N.}~\bibnamefont{Engheta}},
  \bibinfo{journal}{Physical Review B} \textbf{\bibinfo{volume}{86}},
  \bibinfo{pages}{115123} (\bibinfo{year}{2012}).

\bibitem[{\citenamefont{Archambault et~al.}(2012)\citenamefont{Archambault,
  Besbes, and Greffet}}]{Archambault_PRL_2012}
\bibinfo{author}{\bibfnamefont{A.}~\bibnamefont{Archambault}},
  \bibinfo{author}{\bibfnamefont{M.}~\bibnamefont{Besbes}}, \bibnamefont{and}
  \bibinfo{author}{\bibfnamefont{J.-J.} \bibnamefont{Greffet}},
  \bibinfo{journal}{Physical Review Letters} \textbf{\bibinfo{volume}{109}},
  \bibinfo{pages}{097405} (\bibinfo{year}{2012}).

\bibitem[{\citenamefont{Archambault et~al.}(2009)\citenamefont{Archambault,
  Teperik, Marquier, and Greffet}}]{Archambault_PRB_2009}
\bibinfo{author}{\bibfnamefont{A.}~\bibnamefont{Archambault}},
  \bibinfo{author}{\bibfnamefont{T.~V.} \bibnamefont{Teperik}},
  \bibinfo{author}{\bibfnamefont{F.}~\bibnamefont{Marquier}}, \bibnamefont{and}
  \bibinfo{author}{\bibfnamefont{J.~J.} \bibnamefont{Greffet}},
  \bibinfo{journal}{Physical Review B} \textbf{\bibinfo{volume}{79}},
  \bibinfo{pages}{195414} (\bibinfo{year}{2009}).

\bibitem[{\citenamefont{Ashcroft and Mermin}(1976)}]{Ashcroft_Book}
\bibinfo{author}{\bibfnamefont{N.}~\bibnamefont{Ashcroft}} \bibnamefont{and}
  \bibinfo{author}{\bibfnamefont{N.}~\bibnamefont{Mermin}},
  \emph{\bibinfo{title}{Solid State Physics}} (\bibinfo{publisher}{Harcourt
  College Publishers}, \bibinfo{year}{1976}).

\bibitem[{\citenamefont{Henrard et~al.}(1999)\citenamefont{Henrard, Stephan,
  and Colliex}}]{Henrard_SynthMet_1999}
\bibinfo{author}{\bibfnamefont{L.}~\bibnamefont{Henrard}},
  \bibinfo{author}{\bibfnamefont{O.}~\bibnamefont{Stephan}}, \bibnamefont{and}
  \bibinfo{author}{\bibfnamefont{C.}~\bibnamefont{Colliex}},
  \bibinfo{journal}{Synthetic Metals} \textbf{\bibinfo{volume}{103}},
  \bibinfo{pages}{2502} (\bibinfo{year}{1999}).

\bibitem[{\citenamefont{Hoeflich et~al.}(2009)\citenamefont{Hoeflich, Goesele,
  and Christiansen}}]{Hoeflich_PRL_2009}
\bibinfo{author}{\bibfnamefont{K.}~\bibnamefont{Hoeflich}},
  \bibinfo{author}{\bibfnamefont{U.}~\bibnamefont{Goesele}}, \bibnamefont{and}
  \bibinfo{author}{\bibfnamefont{S.}~\bibnamefont{Christiansen}},
  \bibinfo{journal}{Physical Review Letters} \textbf{\bibinfo{volume}{103}},
  \bibinfo{pages}{087404} (\bibinfo{year}{2009}).

\bibitem[{\citenamefont{Henrard et~al.}(2010)\citenamefont{Henrard, VanDenBem,
  Lambin, and Lucas}}]{Henrard_PRL_2010}
\bibinfo{author}{\bibfnamefont{L.}~\bibnamefont{Henrard}},
  \bibinfo{author}{\bibfnamefont{C.}~\bibnamefont{VanDenBem}},
  \bibinfo{author}{\bibfnamefont{P.}~\bibnamefont{Lambin}}, \bibnamefont{and}
  \bibinfo{author}{\bibfnamefont{A.}~\bibnamefont{Lucas}},
  \bibinfo{journal}{Physical Review Letters} \textbf{\bibinfo{volume}{104}},
  \bibinfo{pages}{149701} (\bibinfo{year}{2010}).

\bibitem[{\citenamefont{Hoeflich et~al.}(2010)\citenamefont{Hoeflich, Goesele,
  and Christiansen}}]{Hoeflich_PRL_2010}
\bibinfo{author}{\bibfnamefont{K.}~\bibnamefont{Hoeflich}},
  \bibinfo{author}{\bibfnamefont{U.}~\bibnamefont{Goesele}}, \bibnamefont{and}
  \bibinfo{author}{\bibfnamefont{S.}~\bibnamefont{Christiansen}},
  \bibinfo{journal}{Physical Review Letters} \textbf{\bibinfo{volume}{104}},
  \bibinfo{pages}{149702} (\bibinfo{year}{2010}).

\bibitem[{\citenamefont{Muys}(2012)}]{Muys_OL_2012}
\bibinfo{author}{\bibfnamefont{P.}~\bibnamefont{Muys}},
  \bibinfo{journal}{Optics Letters} \textbf{\bibinfo{volume}{37}},
  \bibinfo{pages}{4928} (\bibinfo{year}{2012}).

\bibitem[{\citenamefont{Berreman}(1963)}]{Berreman_PhysRev_1963}
\bibinfo{author}{\bibfnamefont{D.~W.} \bibnamefont{Berreman}},
  \bibinfo{journal}{Physical Review} \textbf{\bibinfo{volume}{130}},
  \bibinfo{pages}{2193} (\bibinfo{year}{1963}).

\bibitem[{\citenamefont{Ruppin and Englman}(1970)}]{Ruppin_RepProgPhys_1970}
\bibinfo{author}{\bibfnamefont{R.}~\bibnamefont{Ruppin}} \bibnamefont{and}
  \bibinfo{author}{\bibfnamefont{R.}~\bibnamefont{Englman}},
  \bibinfo{journal}{Rep. Prog. Phys.} \textbf{\bibinfo{volume}{33}},
  \bibinfo{pages}{149} (\bibinfo{year}{1970}).

\bibitem[{\citenamefont{Vassant
  et~al.}(2012{\natexlab{a}})\citenamefont{Vassant, Hugonin, Marquier, and
  Greffet}}]{Vassant_OE_2012}
\bibinfo{author}{\bibfnamefont{S.}~\bibnamefont{Vassant}},
  \bibinfo{author}{\bibfnamefont{J.-P.} \bibnamefont{Hugonin}},
  \bibinfo{author}{\bibfnamefont{F.}~\bibnamefont{Marquier}}, \bibnamefont{and}
  \bibinfo{author}{\bibfnamefont{J.-J.} \bibnamefont{Greffet}},
  \bibinfo{journal}{Optics Express} \textbf{\bibinfo{volume}{20}},
  \bibinfo{pages}{23971} (\bibinfo{year}{2012}{\natexlab{a}}).

\bibitem[{\citenamefont{Vassant
  et~al.}(2012{\natexlab{b}})\citenamefont{Vassant, Archambault, Marquier,
  Pardo, Gennser, Cavanna, Pelouard, and Greffet}}]{Vassant_PRL_2012}
\bibinfo{author}{\bibfnamefont{S.}~\bibnamefont{Vassant}},
  \bibinfo{author}{\bibfnamefont{A.}~\bibnamefont{Archambault}},
  \bibinfo{author}{\bibfnamefont{F.}~\bibnamefont{Marquier}},
  \bibinfo{author}{\bibfnamefont{F.}~\bibnamefont{Pardo}},
  \bibinfo{author}{\bibfnamefont{U.}~\bibnamefont{Gennser}},
  \bibinfo{author}{\bibfnamefont{A.}~\bibnamefont{Cavanna}},
  \bibinfo{author}{\bibfnamefont{J.~L.} \bibnamefont{Pelouard}},
  \bibnamefont{and} \bibinfo{author}{\bibfnamefont{J.~J.}
  \bibnamefont{Greffet}}, \bibinfo{journal}{Physical Review Letters}
  \textbf{\bibinfo{volume}{109}}, \bibinfo{pages}{237401}
  (\bibinfo{year}{2012}{\natexlab{b}}).

\end{thebibliography}
\end{document}